\documentclass[prl,superscriptaddress,preprintnumbers,twocolumn,showpacs]{revtex4}
\usepackage{amsmath}
\usepackage{amssymb}
\usepackage{amsthm}
\usepackage{amsfonts}
\usepackage{algorithmic}
\usepackage{enumerate}
\usepackage{latexsym}
\usepackage{graphicx}
\usepackage{bm}
\usepackage{subfigure}
\usepackage{setspace}
\usepackage[dvips]{color}

\begin{document}

\title{Barrier tunneling of the Loop-Nodal Semimetal in the Hyperhoneycomb lattice}
\author{Ji-Huan Guan}
\affiliation{SKLSM, Institute of Semiconductors, Chinese Academy of Sciences, P.O. Box 912, Beijing 100083, China}
\affiliation{School of Physical Sciences, University of Chinese Academy of Sciences, Beijing 101408, China}
\affiliation{Synergetic Innovation Center of Quantum Information and Quantum Physics, University of Science and Technology of China, Hefei, Anhui 230026, China}
\author{Yan-Yang Zhang}
\email{yanyang@semi.ac.cn}
\affiliation{SKLSM, Institute of Semiconductors, Chinese Academy of Sciences, P.O. Box 912, Beijing 100083, China}
\affiliation{School of Microelectronics, University of Chinese Academy of Sciences, Beijing 101408,
China}
\affiliation{Synergetic Innovation Center of Quantum Information and Quantum Physics, University of Science and Technology of China, Hefei, Anhui 230026, China}
\author{Wei-Er Lu}
\affiliation{Microelectronic Instrument and Equipment Research Center, Institute of Microelectronics, Chinese Academy of Sciences, Beijing 100029, China}
\affiliation{Key Laboratory of Microelectronics Devices and Integrated Technology, Institute of Microelectronics, Chinese Academy of Sciences, Beijing 100029, China}
\author{Yang Xia}
\affiliation{Microelectronic Instrument and Equipment Research Center, Institute of Microelectronics, Chinese Academy of Sciences, Beijing 100029, China}
\affiliation{School of Microelectronics, University of Chinese Academy of Sciences, Beijing 101408,
China}
\author{Shu-Shen Li}
\affiliation{SKLSM, Institute of Semiconductors, Chinese Academy of Sciences, P.O. Box 912, Beijing 100083, China}
\affiliation{College of Materials Science and Opto-Electronic Technology, University of Chinese Academy of Sciences, Beijing 101408, China}
\affiliation{Synergetic Innovation Center of Quantum Information and Quantum Physics, University of Science and Technology of China, Hefei, Anhui 230026, China}
\date{\today}

\begin{abstract}

We theoretically investigate the barrier tunneling in the three-dimensional model of the hyperhoneycomb lattice, which is a nodal-line semimetal with a Dirac loop at zero energy. In the presence of a rectangular potential, the scattering amplitudes for different injecting states around the nodal loop are calculated, by using analytical treatments of the effective model, as well as numerical simulations of the tight binding model. In the low energy regime, states with remarkable transmissions are only concentrated in a small range around the loop plane. When the momentum of the injecting electron is coplanar with the nodal loop, nearly perfect transmissions can occur for a large range of injecting azimuthal angles if the potential is not high. For higher potential energies, the transmission shows a resonant oscillation with the potential, but still with peaks being perfect transmissions that do not decay with the potential width. These robust transports of the loop-nodal semimetal can be approximately explained by a momentum dependent Dirac Hamiltonian.
\end{abstract}

\maketitle

\section{I. Introduction}

It was predicted that a relativistic particle can tunnel through an arbitrarily high and wide potential barrier with large probabilities and can even approach perfect for a high barrier, which is called the Klein paradox\cite{Klein1929,Su1993,Calogeracos1999,Dombey1999}.
The test of this prediction has been proved impossible using elementary particles\cite{Greiner1985} but can be experimentally realized in condensed-matter systems, e.g., single layer graphene\cite{Katsnelson2006,Tudorovskiy2012}, where quasiparticles are massless two-dimensional (2D) Dirac fermions. In graphene, unoccupied states in valence band behave as holes, like positrons in particle physics. The matching between electron and hole wave functions at potential barrier boundaries results in the Klein tunneling. Subsequently, Klein tunneling has also been widely studied in bilayer graphene\cite{Katsnelson2006,Tudorovskiy2012,Donck2016}, trilayer graphene\cite{Kumar2012}, multi-layer graphene\cite{Duppen2013} and Weyl semimetals\cite{Hills2017,Yesilyurt2016}, showing various interesting properties.

Recently, a family of trigonally connected three-dimensional (3D) lattices has been put forward\cite{Mullen2015,Mullen2016,Ezawa2016}, termed harmonic honeycomb lattices, which are extensions from the structure of graphene lattice. Among them, the simplest structure is the hyperhoneycomb lattice, which has been discussed intensively in magnetism (i.e., Kitaev model\cite{Kitaev2006})\cite{Mandal2009,Lee2014,Kimchi2014,Hermanns2015,Takayama2015} and superconductivity\cite{Schmidt2016}, and its typical corresponding material is $\beta$-Li$_{2}$IrO$_{3}$\cite{Modic2014}. The tight-binding model of the hyperhoneycomb lattice is supported by the results of first-principles calculations\cite{Mullen2016,Verissimo-Alves2017}.
In the reciprocal space, the conduction and valance bands of graphene touch at two Dirac points, while those of the hyperhoneycomb lattice cross and form a one-dimensional (1D) Dirac loop. Therefore this hyperhoneycomb lattice is a realization of the loop-nodal semimetal. Near the loop, the energy spectrum is linear with respect to the distance from the loop.

In this paper, we study the barrier tunneling for electronic states of the 3D hyperhoneycomb lattice around the nodal loop . Near the nodal loop, the quasiparticles in the hyperhoneycomb lattice follow a momentum dependent Dirac equation, leading to interesting tunneling behaviors, which are analogous to the Klein tunneling in graphene. When the momentum of the injecting electron is coplanar with the loop, scattering can only happen between states with the same pseudo-spin (see below). In the presence of low barrier, almost perfect transmission can happen for a wide range of injecting azimuthal angles, and does not decay with the increasing of the barrier width. For higher barrier energies, the transmission shows a resonant behavior when the barrier height increases, with the resonance peaks back to unity (perfect transmission). These properties reflect the robust transport capabilities of the loop nodal fermions. When the vector of the injecting momentum is deviating from the loop plane, the transmission will decrease to zero soon. In other words, states capable of remarkable tunneling are extremely concentrated in a small range around the plane of the nodal loop. This is not a remarkable downside because we are only interested in low Fermi energies around the nodal loop.

This paper is organized as follows. In Sec. II, we show the model of the hyperhoneycomb lattice based on the tight-binding approximation. In Sec. III, the effective Hamiltonian is used to study the barrier tunneling of quasiparticles near the loop analytically. Sec. IV is devoted to the numerical simulation of tunneling by using the tight binding model. Comparisons with analytical results are also presented. In Sec. V, we summarize results and make conclusions.

\section{II. The Model}
\begin{figure}[htbp]
\includegraphics*[width=0.47\textwidth]{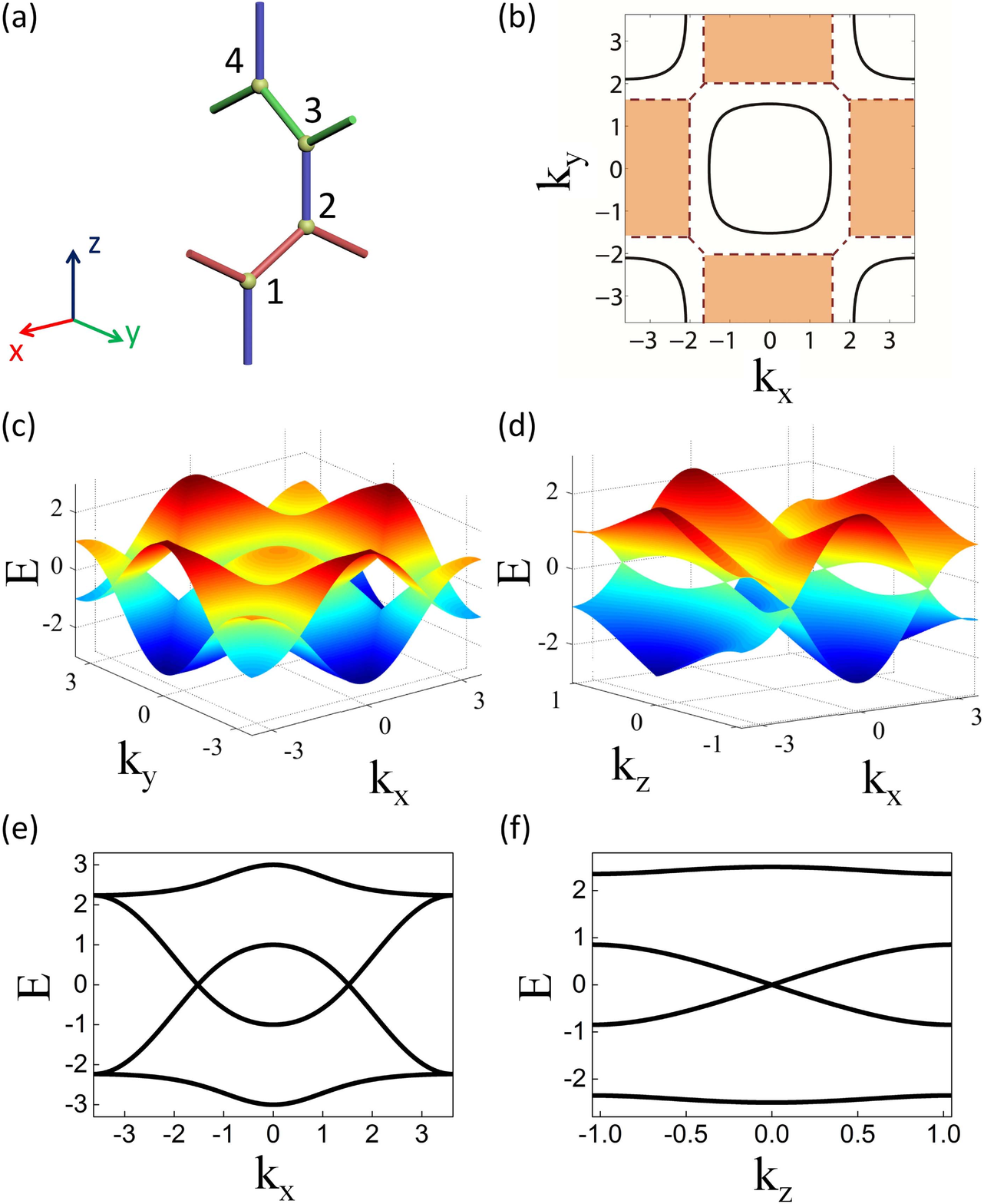}
\caption{(Color online) (a) The unit cell of the hyperhoneycomb lattice, with atoms 1,2,3 and red links on the $x-z$ plane, atoms 2,3,4 and green links on the $y-z$ plane, and blue links along the $z$ direction. (b) The Dirac loop (solid black lines) at zero energy in the $k_{x}-k_{y}$ plane with $k_{z}=0$. The brown (white) region is the top boundary (middle cross section) of the 3D Brillouin zones, which are truncated octahedrons\cite{Mullen2015}. (c) Energy spectra of the lowest two bands in the $k_{x}-k_{y}$ plane, with $k_{z}=0$. (d) Energy spectra of the lowest two bands in the $k_{x}-k_{z}$ plane, with $k_{y}=0$. (e) Band structures along $k_{x}$, with $k_{y}=k_{z}=0$. (f) Band structures along $k_{z}$, with $k_{y}=0$ and $k_{x}=1.522$.}
\label{BANDSTRUCTURE}
\end{figure}

The unit cell of hyperhoneycomb lattice\cite{Mullen2015,Mullen2016,Ezawa2016} is displayed in Fig. \ref{BANDSTRUCTURE}(a), with numbers 1-4 labeling the four atoms in the unit cell. The plane formed by atoms 1,2,3 is perpendicular to that formed by atoms 2,3,4. Here, like in graphene, the chemical bonds of any three nearest neighboring atoms make an angle of $120^{\circ}$, which is called the trigonal connectivity. The corresponding first Brillouin zone (BZ) is
a truncated octahedron\cite{Mullen2015}. The general form of the tight-binding Hamiltonian in the $k$ space can be expressed as
$\sum_{\bm{k}}\sum_{\alpha\beta}H_{\alpha\beta}(\bm{k})c^{\dagger}_{\bm{k}\alpha}c_{\bm{k}\beta}$, where $\alpha,\beta\in\{1,2,3,4\}$ label the atomic orbitals in the unit cell. In the presence of only nearest hoppings $t$ along these bonds, the matrix elements read
\begin{equation}
H_{\alpha\beta}(\bm{k})=\left(
\begin{array}{cccc}
0 & f_{x} & 0 & f_{z}^{*}\\
f_{x}^{*} & 0 & f_{z} & 0\\
0 & f_{z}^{*} & 0 & f_{y}\\
f_{z} & 0 & f_{y}^{*} & 0
\end{array}%
\right),  \label{Hamiltonian}
\end{equation}%
where
\begin{eqnarray}
&&f_{x}=\left(t e^{i \frac{\sqrt{3}}{2} k_{x} a}+t e^{-i \frac{\sqrt{3}}{2} k_{x} a}  \right) e^{i \frac{k_{z}}{2} a},\nonumber\\
&&f_{y}=\left(t e^{i \frac{\sqrt{3}}{2} k_{y} a}+t e^{-i \frac{\sqrt{3}}{2} k_{y} a}  \right) e^{i \frac{k_{z}}{2} a},\nonumber\\
&&f_{z}=e^{i k_{z} a},\nonumber
\end{eqnarray}
and $a$ is the interatomic distance. For convenience, we set $a=1$ and $t=1$ as length and energy units respectively throughout this paper.
Solving the energy eigenvalue equation $H\Psi=E\Psi$ for zero energy, one obtains $k_{z}=0$ and the relation\cite{Ezawa2016}
\begin{equation}
4\cos(\sqrt{3}k_{x}a/2)\cos(\sqrt{3}k_{y}a/2)=1, \label{loop}
\end{equation}
the solution of which gives the nodal loop $\bm{k}_{0}=(k_{0x},k_{0y},0)$ in the plane $k_z=0$ and centered around $\Gamma=(0,0,0)$, as depicted in Fig. \ref{BANDSTRUCTURE}(b). Fig. \ref{BANDSTRUCTURE}(c) and (d) show the lowest two bands around the loop in $k_{x}-k_{y}$ space and in $k_{x}-k_{z}$ space respectively, where one can see the loop line given by Eq. (\ref{loop}). The complete four bands solved from Hamiltonian (\ref{Hamiltonian}) in given directions are plotted in Fig. \ref{BANDSTRUCTURE}(e)(f), with $k_{y}=k_{z}=0$, and with $k_{y}=0$, $k_{x}=1.522$, respectively. In the following, we will investigate the tunneling properties of states around the nodal loop.

\section{III. Analytical Results from the Effective Model}
In order to focus on the quasiparticle properties near the nodal loop, we adopt a 2$\times$2 effective Hamiltonian which is projected from Hamiltonian (\ref{Hamiltonian}) to the two-component subspace that represents the
lowest energy bands around the nodal loop. The effective Hamiltonian can be written as ($\hbar\!=\!1$ throughout this manuscript)\cite{Mullen2015}
\begin{equation}
H_{\mathrm{eff}}(\phi,\bm{q})=-\left[v_{x}(\phi)q_{x}+v_{y}(\phi)q_{y}\right]\sigma_{x}+v_{z}(\phi)q_{z}\sigma_{z},
\label{effequation}
\end{equation}
where $\sigma_{i}$ are pseudo-spin Pauli matrices acting on this subspace. The momentum
\begin{equation}
\bm{q}=\bm{k}(\phi)-\bm{k}_{0}(\phi) \label{EqDefineQ}
\end{equation}
is measured from a nodal point on the nodal loop $\bm{k}_{0}(\phi)=\big(k_{0x},k_{0y},0\big)$ [$k_{0x}$ and $k_{0y}$ satisfy Eq. (\ref{loop})]. The azimuthal angle $\phi$ and the polar angle $\theta$ are defined in the usual way as
\begin{equation} \label{EqDefineAngles}
\tan \phi=\frac{k_{0y}}{k_{0x}},  \qquad \cos \theta=\frac{k_{z}}{k}.
\end{equation}
Since the loop is within the $k_z=0$ plane, the out-of-plane component is simply $q_z=k_z$.

\begin{figure}[htbp]
\includegraphics*[width=0.35\textwidth]{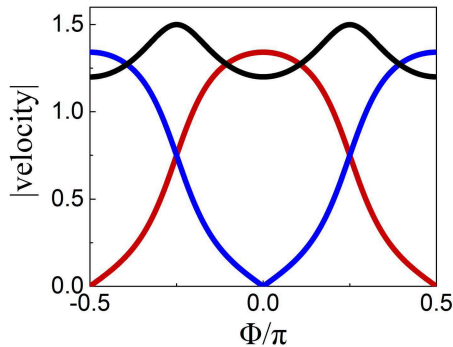}
\caption{(Color online) Velocity of the quasiparticles at the loop as a function of the angle $\phi$, where the red is $v_{x}$, the blue is $v_{y}$ and the black is $v_{z}$.}
\label{velocity}
\end{figure}

\begin{figure}[htbp]
\includegraphics*[width=0.40\textwidth]{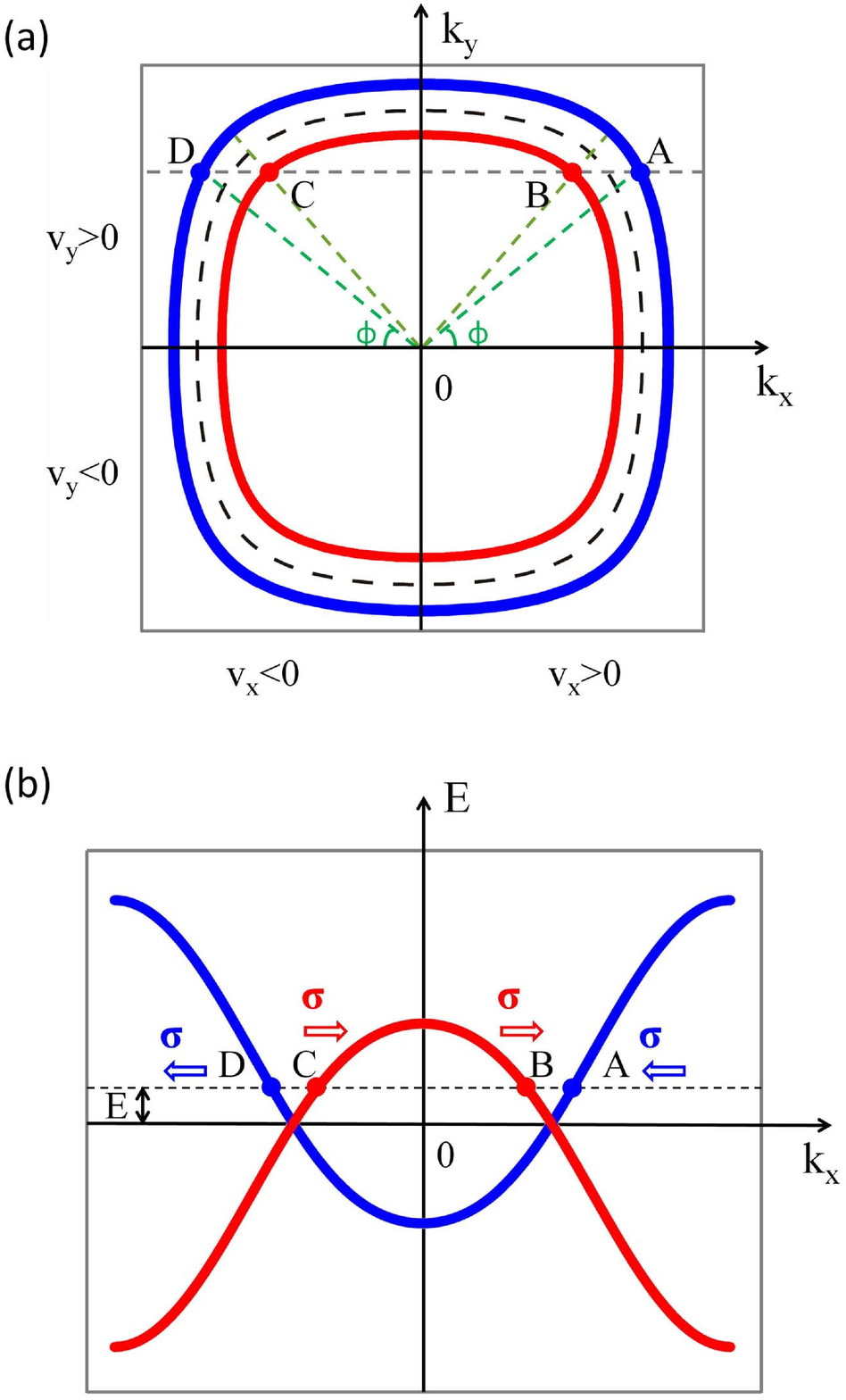}
\caption{(Color online) (a) The cross section of the band structure (solid circles) in Fig. \ref{BANDSTRUCTURE}(c) at a given Fermi energy $E>0$. The dashed circle denotes the loop position at zero energy. (b) The cross section of the band structure in Fig. \ref{BANDSTRUCTURE}(c) at a given $k_y$. The pseudo-spin for the red (blue) branch of states is $\sigma_x=+1$ ($\sigma_x=-1$).}
\label{pseudospin}
\end{figure}

Due to the 1D nature of the nodal loop, notice that for an unambiguous definition of $\bm{q}$ in Eq. (\ref{EqDefineQ}), the reference point $\bm{k}_0$ is $\bm{k}$ dependent and is chosen to be the nodal point $\bm{k}_0$ that has the same azimuthal angle $\phi$ with $\bm{k}$ [see Fig.\ref{pseudospin} (a)]. Three components of the Fermi velocity, $v_{x},v_{y},v_{z}$ in Eq. (\ref{effequation}) are also $\bm{k}$ dependent, and their dependence on the azimuthal angle $\phi$ is plotted in Fig. \ref{velocity}. For a globally smooth definition of (\ref{effequation}) throughout the loop region, it is convenient to define
\begin{equation}\label{EqDefineVSign}
v_x(\bm{k}_0)\left\{
\begin{array}{lr}
\!>\!0,  &  k_{0x}\!>\!0,   \\
\!<\!0,  &  k_{0x}\!<\!0,
\end{array}
\right.\quad
v_y(\bm{k}_0)\left\{
\begin{array}{lr}
\!>\!0,  &  k_{0y}\!>\!0,   \\
\!<\!0,  &  k_{0y}\!<\!0,
\end{array}
\right.
\end{equation}
as shown in Fig. \ref{pseudospin}(a). For any state point $\bm{q}$ on these two bands around the loop, in the linear dispersion approximation, the velocity components $v_i$ in Eq. (\ref{effequation}) can be considered to be the polar angle $\theta$ independent\cite{Mullen2015}. In a nutshell, such a complicated $\bm{k}$ dependence of the effective Hamiltonian (\ref{effequation}) makes rigorous treatments of its scattering problems much more complicated than those in the Dirac systems, where there is only a fixed nodal point. Therefore, further reasonable approximations should be adopted in the following analytical calculations.

Let us start from the simplest case, $k_{z}=q_{z}=0$, i.e., the states coplanar with the nodal loop.
With the substitutions $q_{x(y)} \rightarrow -i\partial_{x(y)}$ and $q_z \rightarrow 0$, and the plane wave ansatz for the solution
$\Psi(x,y)= e^{i(q_x x + q_y y)}\psi$,
the eigenvalues of Eq. (\ref{effequation}) for $k_{z}=0$ are
\begin{equation}
E_{1,2}(\phi,\bm{q})=\mp (v_{x}q_{x} + v_{y}q_{y}). \label{E1}\\
\end{equation}
The corresponding eigenstates are (up to a normalization factor)
\begin{equation}
\Psi_{1,2}(\phi,\bm{q})=\left( \begin{array}{c} \pm 1 \\ 1 \end{array} \right)e^{iq_{x}x+iq_{y}y}.\label{EqEffEigenstates}\\
\end{equation}
With $\phi$ varying from $0$ to $2\pi$ with the conventions in Eq. (\ref{EqDefineVSign}), these two branches of eigenvalues in Eq. (\ref{E1}) cross at $E=0$ and reproduce the loop-nodal structure, as illustrated in Fig. \ref{pseudospin}.
In this sense, we call the set of eigenvalues $E_{n}(\phi,\bm{q})$ with a definite subscript $n$ a definite ``branch''.
In Fig. \ref{pseudospin}, each branch is plotted with the same color (red or blue). In the absence of the $q_z$ term in Eq. (\ref{effequation}),
now an eigenstate of the effective Hamiltonian operator $H_{\mathrm{eff}}=-\left[v_{x}(\phi)q_{x}+v_{y}(\phi)q_{y}\right]\sigma_{x}$ is also that of the pseudo-spin operator $\sigma_x$. Namely, a definite branch of eigenstates can be labelled by its pseudo-spin eigenvalue ($\sigma_x=+1$ or $\sigma_x=-1$), as illustrated in Fig. \ref{pseudospin}. This pseudo-spin plays an important role in discussing the in-plane scattering, as will be seen soon.

Now consider a cubic barrier with a pseudo-spin independent potential, i.e., in the form of $\sigma_0 V(\bm{x})$,
with
\begin{equation}\label{Barrier}
V(\bm{x})=\left\{
\begin{array}{lr}
V_{0},  &  0<x<d,   \\
0,      &  x<0, x>d,
\end{array}
\right.
\end{equation}
and extending infinitely along the $y$ and $z$ directions. We assume that an incident electron with energy $E=v_{x}q_{x} + v_{y}q_{y}>0$ (remember $q_z=0$ so far) is injecting towards this barrier.
Since the potential (\ref{Barrier}) does not break the translation symmetry in $y$ and $z$ directions, $k_y$ and $k_z$ are conserved throughout the tunneling process.

As mentioned above, an analytically rigorous calculation of the scattering problem for Model (\ref{effequation}) is technically difficult. However, as will be seen in the next section, even after some very rough approximations, the results still offer good insights qualitatively, and even provide excellent agreements with the numerical simulation for the full tight binding model quantitatively (in most of the parameter space). This fact reflects the robust transport properties of this nodal loop semimetal.

As illustrated in Fig. \ref{pseudospin}, for a given Fermi energy $E$, $k_y$ and $k_z$ ($k_z=0$ here), there are four propagating states $A,B,C,D$ on two branches, where $A,C$ are the right-moving states and $B,D$ the left-moving states. To be spesific, we take $A$ as the incident state. Then the rest three states (with identical $k_y$, $k_z$ and $E$) are those it can be scattered to: $B$ and $D$ serve as the reflection states, and $C$ serves as the transmission state.
To simplify the analytical calculations, we will treat these different scattering processes \emph{separately} in the following.

Firstly we take $B$ as the only state onto which the electron can be scattered, i.e., the process $A\rightarrow B$. Note that $A$ and $B$ correspond to different $\phi$ (but the difference is slight for small energy $E$ around the loop), and therefore correspond to different (but close) reference points $\bm{k}_0$ for their effective Hamiltonians respectively, as can be seen in Fig. \ref{pseudospin}(a). This makes the problem difficult to solve. To circumvent this obstacle and to concentrate on the dominating physics, we simply ignore this slight difference and introduce the approximation that they belong to the same reference point on the loop. On the other hand, we still require the conservation of momentum in the $y$ direction, i.e., identical $q_{y}=0$ for both states. This approximation of unification to a single reference point on the nodal loop is reasonable for the low energy regime. We further assume this to be applicable to the states inside the barrier (with an energy shift $V_0$) as well, as long as the barrier $V_0$ is low enough. With the full tight binding simulations in section IV, the validity of these approximations will be supported, and how the results deviate for high energy $E$ and high barrier $V$ will also be shown.

Now, the wave functions in three regions are approximately given by
\begin{equation*}
\Psi(x)\!=\!
\left\{
\begin{array}{lr}
\!\left(\!\begin{array}{c} -1 \\ 1 \end{array}\!\right)\!e^{iq_{x}x}+r\!\left(\!\begin{array}{c} 1 \\ 1 \end{array}\!\right)\!e^{-iq_{x}x}, & x\!<\!0,  \\[4mm]
 b\!\left(\!\begin{array}{c} -1 \\ 1 \end{array}\!\right)\!e^{ip_{x}x}+c\!\left(\!\begin{array}{c} 1 \\ 1 \end{array}\!\right)\!e^{-ip_{x}x}, & 0\!<\!x\!<\!d,   \\[4mm]
t\!\left(\!\begin{array}{c} -1 \\ 1 \end{array}\!\right)\!e^{iq_{x}x}, & x\!>\!d,
\end{array}
\right.
\end{equation*}
where $q_{y}$ has been set to zero, $r, t, b, c$ are coefficients of wave functions, $r$ ($t$) is the reflection (transmission) coefficient, and $q_{x}$ and $p_{x}$ are the wave vectors measured from the point $k_{0x}$ outside and inside the barrier region respectively.

Using the continuity of the wave functions at $x=0$ and $x=d$, one can determine the transmission coefficient
\begin{equation}
t=e^{-iq_{x}d+ip_{x}d},
\end{equation}
and the transmission probability
\begin{equation}
T=|t|^{2}=1. \label{EqPerfectTransmission}
\end{equation}

This result is not surprising since it is just the 1D version of Klein tunneling:
with $q_z=0$ and $q_y=0$, Eq. (\ref{effequation}) is just like the Hamiltonian of a 1D massless Dirac fermion, or, the normal incident channel of the 2D massless Dirac fermion\cite{Katsnelson2006,Neto2009,Peres2010,Tudorovskiy2012}. Therefore electrons can penetrate through the potential barrier without any backscattering, as long as the state is not far from the nodal loop where the above approximations hold. This perfect tunneling has nothing to to with the barrier height nor the angle $\phi$, which is different from the 2D case in graphene\cite{Katsnelson2006}. Intuitively, this perfect transmission can be understood as a conservation of the pseudo-spin\cite{Tudorovskiy2012} associated with the Dirac-like Hamiltonian (\ref{effequation}): States $A$ and $B$ belong to branches with different pseudo-spins [see Fig. \ref{pseudospin}],
so scattering between them is forbidden.

Secondly, we discuss the scattering process $A\rightarrow D$. In the reciprocal space, states $A$ and $D$ are well separated and symmetric with respect to the $E$ axis. In the effective Hamiltonians (\ref{effequation}) and their eigenstates (\ref{EqEffEigenstates}), the wavevectors $\bm{q}$ for them are defined with respect to different nodal points $\bm{K}_0$ and $-\bm{K}_0$ respectively, i.e., $\bm{q}_{A(D)}=\bm{k}\mp \bm{K}_0$, where $\bm{K}_0=(k_{0x},k_{0y},0)$.
To unify the notations in a single equation, we can simply express these eigenstates as $\Psi_{A(D)}(\bm{k}\mp \bm{K}_0)$ instead of $\Psi_{A(D)}(\bm{q}_{A(D)})$.
Since $k_y$ is conserved, it only represents an irrelevant constant phase factor and can be neglected in the calculations.
Now, the wave functions in three regions are approximately given by
\begin{equation*}
\begin{array}{l}
\Psi(x)=\\[3mm]
\!\left\{\!
\begin{array}{lr}
\left[ \!\left(\!\begin{array}{c} -1 \\ 1 \end{array}\!\right)\!e^{ik_{x}x}+r\!\left(\!\begin{array}{c} -1 \\ 1 \end{array}\!\right)\!e^{-ik_{x}x}\!\right]\!e^{-ik_{0x}x}, \quad x\!<\!0,  \\[4mm]
\left[b\!\left(\!\begin{array}{c} -1 \\ 1 \end{array}\!\right)\!e^{ik_{x}^{\prime}x}+c\!\left(\!\begin{array}{c} -1 \\ 1 \end{array}\!\right)\!e^{-ik_{x}^{\prime}x}\!\right]\!e^{-ik_{0x}x}, \quad 0\!<\!x\!<\!d,   \\[4mm]
\left[t\!\left(\!\begin{array}{c} -1 \\ 1 \end{array}\!\right)\!e^{ik_{x}x}\!\right]\!e^{-ik_{0x}x},\quad x\!>\!d,
\end{array}
\right.
\end{array}
\end{equation*}
where $k_{x}$, $k_{x}^{\prime}$ and $k_{0x}$ are positive. The second terms in the first two equations represent the left-moving state $D$, and $k_{x}$, $k_{x}^{\prime}$ are the wave vectors in the $x$ direction outside and inside the barrier region respectively, which satisfy
\begin{equation} \label{Eqkx}
\begin{split}
E&=v_{x}(k_{x}-k_{0x})+v_{y}q_{y},  \\
E-V_{0}&=v_{x}(k_{x}^{\prime}-k_{0x})+v_{y}p_{y},
\end{split}
\end{equation}
where $q_{y}\approx p_{y}$ for the momentum conversation and the proper approximation inside the barrier.
By matching the wave functions at the barrier boundaries, the resulting transmission is given by
\begin{equation}
T\!=\!|t|^{2}\!=\! \frac{4k_{x}^{2}k_{x}^{\prime2}}{4k_{x}^{2}k_{x}^{\prime2}\cos^{2}(k_{x}^{\prime}d)+(k_{x}^{2}\!+\!k_{x}^{\prime2})^{2}\sin^{2}(k_{x}^{\prime}d)}. \label{T2}
\end{equation}

From Eqs. (\ref{Eqkx}-\ref{T2}), some useful conclusions can be made. If $V_{0}$ is small, the values of $k_{x}$ and $k_{x}^{\prime}$ do not differ much, and therefore the transmission probability in Eq. (\ref{T2}) can still approach unity. As can be seen from Eq. (\ref{Eqkx}), the difference $\big|k_{x}-k_{x}^{\prime}\big|$ is proportional to $|\frac{V_{0}}{v_{x}}|$. When the incident angle $\phi$ increases, $|v_{x}|$ decreases (see the red line in Fig. \ref{velocity}) and the difference between $k_{x}$ and $k_{x}^{\prime}$ becomes larger, which causes the imperfect transmission. Nevertheless, there still exists the resonance condition for the perfect tunneling $T=1$, when $k_{x}^{\prime}d=n\pi$, where $n$ is an integer. By using Eq. (\ref{Eqkx}), this resonance condition can be expressed as
\begin{equation}
V_{0}=E+v_{x}k_{x0}-v_{y}p_{y}-\frac{\pi v_{x}}{d}n, \quad n\in \mathrm{integer}.\label{EqResonantCondition}
\end{equation}
The resonant frequency is related to the barrier width $d$ and the $x$-component of velocity $v_{x}$, where $v_{x}$ is a function of the incident angle $\phi$ (Fig. \ref{velocity}). For a given incident angle and a fixed barrier width, the resonant period keeps constant with the increasing of the barrier height $V_{0}$. This feature will be seen from full tight binding numerical simulations in the next section.

Thirdly, we turn to the scattering process of transmission from state $A$ to state $C$.
The evaluating of this transmission is even more tricky. For example, besides these two right-going states, at least one reflection state ($B$ or $D$) should also be accounted in, which will involve too many undetermined coefficients. We adopt the following simple approximations to make this solvable.
Still, state $A$ is the incident channel and state $D$ serves as the only reflection channel. Inside the barrier, right-moving state is state $A$ or state $C$. In the third region $x>d$, state $C$ is the only transmission channel. Since the states $C$ and $D$ are close in the reciprocal space, we make a further approximation that the wavevector of state $C$ is the same as that of $D$. Now the wave functions in three regions are roughly written as
\begin{equation*}
\begin{array}{l}
\Psi(x)=\\[2mm]
\!\left\{\!
\begin{array}{lr}
\!\left[\!\left(\!\begin{array}{c} -1 \\ 1 \end{array}\!\right)\!e^{ik_{x}x}+r\!\left(\!\begin{array}{c} -1 \\ 1 \end{array}\!\right)\!e^{-ik_{x}x}\!\right]\!e^{-ik_{0x}x}, & x\!<\!0,  \\[4mm]
\!\left[b\!\left(\!\begin{array}{c} \mp1 \\ 1 \end{array}\!\right)\!e^{\pm ik_{x}^{\prime}x}+c\!\left(\!\begin{array}{c} -1 \\ 1 \end{array}\!\right)\!e^{-ik_{x}^{\prime}x}\!\right]\!e^{-ik_{0x}x}, & 0\!<\!x\!<\!d,   \\[4mm]
\!\left[t\!\left(\!\begin{array}{c} 1 \\ 1 \end{array}\!\right)\!e^{-ik_{x}x}\!\right]\!e^{-ik_{0x}x}, & x\!>\!d,
\end{array}
\right.
\end{array}
\end{equation*}
where the sign $\mp$ in the column vector denotes state $A$ and state $C$ respectively. Solving the equations at the barrier boundary, we get the transmission $T=0$ for both ``$-$'' and ``$+$'' cases, which means there is no transmission from state $A$ to state $C$. This result will also be supported by the numerical simulations in the next section.

The above tunneling processes for an in-plane ($k_z=0$) injection can be summarized as follows. As illustrated in Fig. \ref{pseudospin}(b), the nodal loop is the crossing of two branches of eigenstates at $E=0$, a red one and a blue one. As discussed above, these two branches can be characterized by well-defined pseudo-spin orientations: $\sigma_x=1$ ($\sigma_x=-1$) for the red (blue) branch. At a finite incident energy $E$, the remarkable scattering can only happen between states possessing the same pseudo-spin, for example, $A\rightarrow A$ (transmission), and $A \rightarrow D$ (reflection). The scattering processes between different pseudo-spins (e.g., transmission $A\rightarrow C$ and reflection $A\rightarrow B$) are prohibited.
This conservation of the pseudo-spin relies on two facts: (i) the eigenstates of the Hamiltonian are also the eigenstates of the one of the pseudo-spin operator $\sigma_x$ due to the absence of $\sigma_z$ term; (ii) the barrier potential (\ref{Barrier}) is independent of the pseudo-spin, and thus does not flip the pseudo-spin.
Similar pseudo-spins related scattering can also be seen in Dirac type systems\cite{Katsnelson2006,Hills2017}.

The discussions above are limited within the case of $k_{z}=0$. Here we will take finite $k_z=q_z$ into account.
Now the eigenvalues of Eq. (\ref{effequation}) are
\begin{equation}
E_{1(2)}=\mp\sqrt{(v_{x}q_{x}+v_{y}q_{y})^{2}+(v_{z}q_{z})^{2}}. \label{E1p}
\end{equation}
The corresponding wave functions are
\begin{equation}
\Psi_{1(2)}=\left( \begin{array}{c} \alpha \pm \beta \\ 1 \end{array} \right)e^{iq_{x}x+iq_{y}y+iq_{z}z},
\end{equation}
where
\begin{eqnarray}
\alpha &\!=\!&\frac{-v_{z}q_{z}}{v_{x}q_{x}+v_{y}q_{y}},\nonumber\\
\beta &\!=\!&\frac{\sqrt{(v_{x}q_{x}+v_{y}q_{y})^{2}+(v_{z}q_{z})^{2}}}{v_{x}q_{x}+v_{y}q_{y}}.\nonumber
\end{eqnarray}

\begin{figure}[htbp]
\includegraphics*[width=0.47\textwidth]{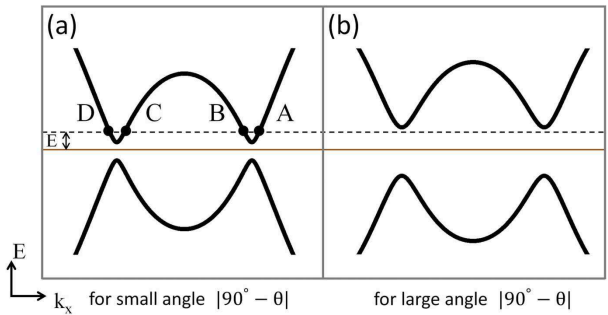}
\caption{(Color online) The lowest two bands for small (left) and large (right) angle $|90^{\circ}-\theta|$ (or value $k_{z}$). The brown horizonal line shows the position of zero energy, and the dashed line denotes the Fermi energy, which crosses the bands in the left at points $A,B,C,D$.}
\label{bands-theta}
\end{figure}

In the presence of the massive term $v_{z}(\phi)q_{z}\sigma_{z}$ in the Hamiltonian, the 2D bands $E_{1(2)}(k_x,k_y,k_z\!=\!\mathrm{const})$ open a gap, which is proportional to $\big|q_z=k_z\big|$,
as illustrated in Fig. \ref{bands-theta}. This leads to two consequences.

The first consequence is quite direct: if we are only interested in the low energy region, then $|k_z|$ (or $|\theta-90^{\circ}|$) should be limited to be small. Otherwise, the Fermi energy $E$ will lie in the gap and there will be no states available, as illustrated in Fig. \ref{bands-theta}(b). The second consequence is more subtle: the presence of the finite mass makes the orientation of the pseudo-spin $\big(\sigma_x,\sigma_z\big)$ precess with $\bm{q}$ for each branch of eigenstates. As a result, remarkable scattering can happen between two branches, since their pseudo-spins are not exactly antiparallel any more\cite{Tudorovskiy2012,Neto2009,Peres2010}. This is similar to the appearance of backscattering in graphene with a finite gap (massive Dirac fermion)\cite{Katsnelson2006}.

As indicated in Fig. \ref{bands-theta}(a), we still use letters $A-D$ to label those four states with the same $k_y$, $k_z$ and $E$. For simplicity, we only calculate the first case where $A$ and $B$ serve as the propagating states, when $q_y=0$. The wave functions in three regions are approximately given by
\begin{equation*}
\begin{array}{l}
\Psi(x)=\\[2mm]
\!\left\{\!\!
\begin{array}{lr}
\!\left[ \!\left(\!\begin{array}{c} \alpha_{1}\!-\!\beta_{1} \\ 1 \end{array}\!\right)\!\!e^{iq_{x}x}\!+\!r\!\left(\!\begin{array}{c} \beta_{1}\!-\!\alpha_{1}\\ 1 \end{array}\!\right)\!e^{-iq_{x}x}\!\right]\!e^{iq_{z}z},\! & x\!<\!0,  \\[4mm]
\!\left[\!b\!\left(\!\begin{array}{c} \alpha_{2}\!-\!\beta_{2} \\ 1 \end{array}\!\right)\!e^{ip_{x}x}\!+\!c\!\left(\!\begin{array}{c} \beta_{2}\!-\!\alpha_{2} \\ 1 \end{array}\!\right)\!e^{-ip_{x}x}\!\right]\!e^{iq_{z}z},\! & 0\!<\!x\!<\!d,   \\[4mm]
t\!\left(\!\begin{array}{c} \alpha_{1}\!-\!\beta_{1} \\ 1 \end{array}\!\right)\!e^{iq_{x}x+iq_{z}z},\! & x\!>\!d,
\end{array}
\right.
\end{array}
\end{equation*}
$q_{y}=0$ and $q_{z}=k_{z}$ are conserved in the process of tunneling, $q_{x}$, $p_{x}$ are defined as
\begin{eqnarray}
q_{x}&=&\frac{\sqrt{E^{2}-(v_{z}k_{z})^{2}}}{v_{x}},\nonumber\\
p_{x}&=&\frac{\sqrt{(E-V_{0})^{2}-(v_{z}k_{z})^{2}}}{v_{x}},\nonumber
\end{eqnarray}
and
\begin{eqnarray}
\alpha_{1}&=&\frac{-v_{z}k_{z}}{\sqrt{E^{2}-(v_{z}k_{z})^{2}}},\nonumber\\
\beta_{1}&=&\frac{E}{\sqrt{E^{2}-(v_{z}k_{z})^{2}}},\nonumber\\
\alpha_{2}&=&\frac{-v_{z}k_{z}}{\sqrt{(E-V_{0})^{2}-(v_{z}k_{z})^{2}}},\nonumber\\
\beta_{2}&=&\frac{E-V_{0}}{\sqrt{(E-V_{0})^{2}-(v_{z}k_{z})^{2}}}.\nonumber
\end{eqnarray}
Using the method of transfer matrix\cite{Hills2017}, we obtain the transmission probability
\begin{eqnarray}\label{EqTFinitek}
\begin{split}
T=\frac{16(\alpha_1-\beta_1)^{2}(\alpha_2-\beta_2)^{2}}
{\!\gamma_{-}^{4}\sin^{2}(2p_{x}d)+[\gamma_{+}^{2}\!-\!\gamma_{-}^{2}\!\cos(2p_{x}d)]^{2}},
\end{split}
\end{eqnarray}
where
\begin{eqnarray}
&\gamma_{+}=(\alpha_{1}\!-\!\beta_{1})+(\alpha_{2}\!-\!\beta_{2}),\nonumber\\
&\gamma_{-}=(\alpha_{1}\!-\!\beta_{1})-(\alpha_{2}\!-\!\beta_{2}).\nonumber
\end{eqnarray}
When $k_{z}=0$, we have $\alpha_{1}=\alpha_{2}=0$, $\beta_{1}=\pm1$, $\beta_{2}=\pm1$, and $T=1$, which is the case discussed in Eq. (\ref{EqPerfectTransmission}). When $k_{z}\neq0$, the values $\alpha_{1}\neq\alpha_{2}$, $\beta_{1}\neq\pm1$, $\beta_{2}\neq\pm1$, and the transmission probability $T$ declines. Nevertheless, there still exists the resonance condition of perfect tunneling $T=1$:
\begin{equation}
2p_{x}d=n\pi, \label{EqResonanceFinitekz}
\end{equation}
which we will see in the following numerical results (resonance peaks in Fig. \ref{Ttheta}).

\section{IV. Numerical Results from the Tight Binding Model}

\begin{figure}[htbp]
\includegraphics*[width=0.47\textwidth]{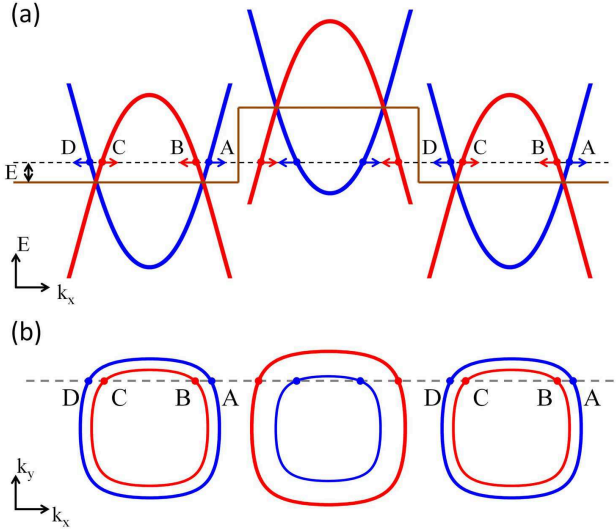}
\caption{(Color online) Schematic of electronic states in three regions of tunneling, with $k_{z}=0$, (a) the ``side view'' (the cross section at a given $k_y$); (b) the ``top view'' (the cross section at the Fermi energy $E$).  The states with pseudo-spin $\sigma_x=+1$ ($\sigma_x=-1$) are plotted in red (blue) color. In (a), brown solid lines show the potential profiles and the horizonal dashed line denotes the Fermi energy ($E<V_{0}$). Small arrows indicate the directions of Fermi velocities. }
\label{electronstates}
\end{figure}

In this section, we will perform numerical simulations on the quantum transport through the barrier based on the full tight-binding model, Eq. (\ref{Hamiltonian}). We still consider that a cubic potential described by Eq. (\ref{Barrier}) is embedded in an infinite 3D hyperhoneycomb lattice, and that the electron is injecting from the left of the barrier (thus with $v_x>0$). As before, such a potential only permits scattering among states with identical $k_y$ and $k_z$ (e.g., states $A$ to $D$ illustrated in Fig. \ref{electronstates}), due to the momentum conservation in these two directions. In other words, the original barrier transmission problem in 3D can be safely decomposed into that of independent 1D transmission models, where each 1D model is characterized by a pair of parameters $(k_y,k_z)$\cite{Kumar2012}. By using the well-developed numerical mode-matching methods for tight-binding models\cite{Ando1991,Khomyakov2005}, we can directly calculate the mode-resolved scattering amplitude $T_{mn}$ from state $n$ to $m$.

\begin{figure}[htbp]
\includegraphics*[width=0.47\textwidth]{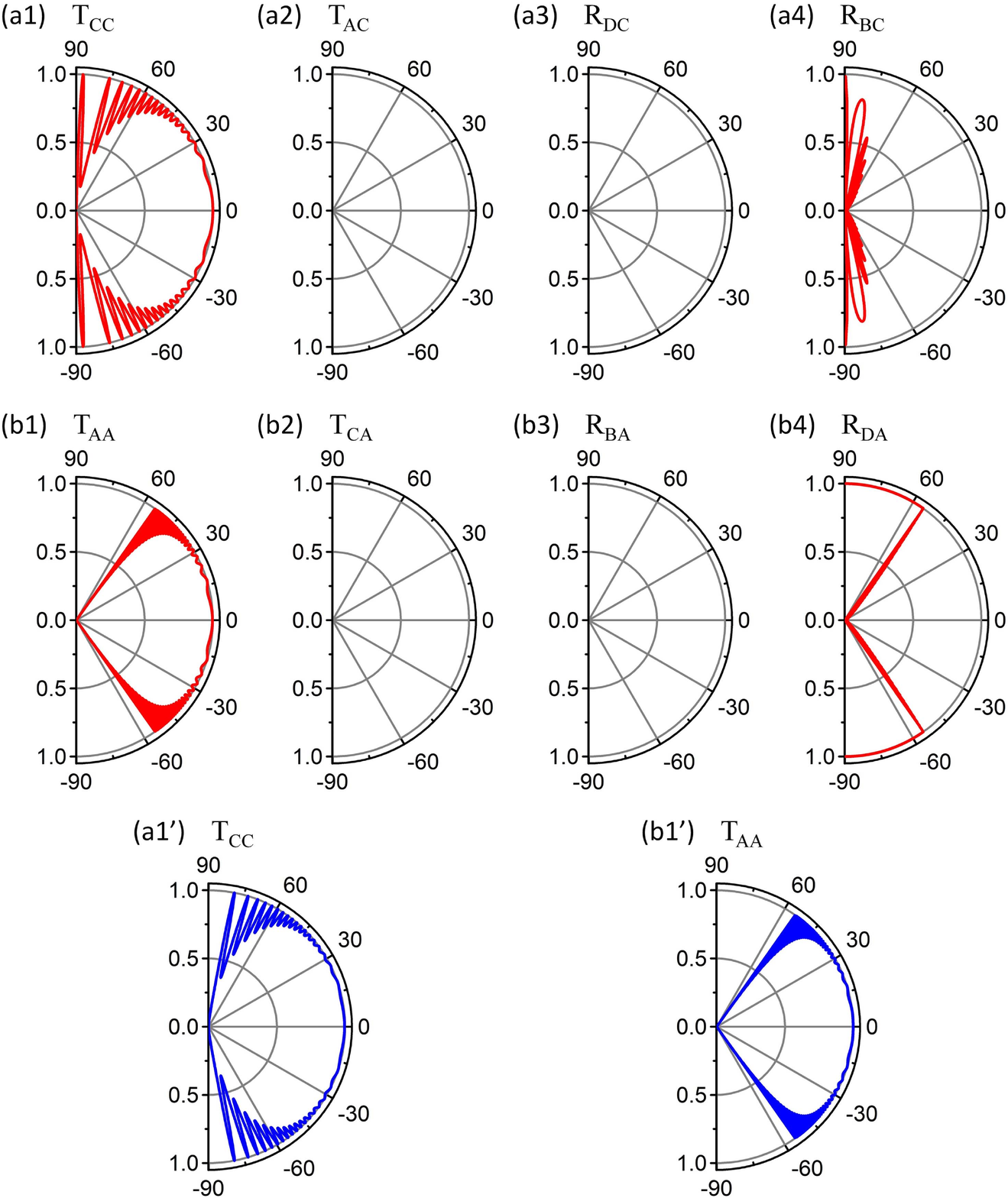}
\caption{(Color online) Transmission ($T$) and reflection ($R$) probabilities as a function of the azimuthal angle $\phi$, for the in-plane incidence $\theta=90^{\circ}$. The incident Fermi energy $E=0.1$, the barrier potential $V_{0}=0.2$, and the barrier width $d=100\sqrt{3}/2$. (a) and (b) correspond to incidence from state $C$ and $A$, respectively, calculated from the tight binding model. (a1') and (b1') are transmissions calculated from the effective model.}
\label{probability}
\end{figure}

Let us still start from the simple case $k_z=0$. In Fig. \ref{probability}, we plot the state resolved scattering amplitudes $T_{mn}$ through the barrier with height $V_0=0.2$, as functions of the incident azimuthal angle $\phi = \arctan(k_y/k_x)$, with a constant incident polar angle
$\theta \!=\!90^{\circ}$ (i.e., $k_z\! = \!0$) and a constant incident energy $E=0.1$. The first (second) row is the results for an incident state $C$ ($A$) on the red (blue) branch [see Fig. \ref{electronstates}], respectively. Let us first concentrate on the first row, $T_{mC}$ with the incident state $C$. The most remarkable feature is that both inter-branch scattering processes, the transmission $T_{AC}$ [Fig. \ref{probability}(a2)] and the reflection $T_{DC}$ [Fig. \ref{probability}(a3)], are identically zero. These results are consistent with the first and the third cases of discussions in last section. As a consequence, there only exist the intra-branch reflection $T_{BC}$ and the intra-branch transmission $T_{CC}$, corresponding to the second case in last section. For comparison, we also plot the transmission from analytical calculation, Eq. (\ref{T2}), in Fig. \ref{probability}(a1'). One can see good agreement between (a1) and (a1') in Fig. \ref{probability}. The disagreement at large incident angle can be attributed to the deviation of the analytical approximation. Similar agreements with analytical results can also be seen for an injecting electron with state $A$, as shown in Fig. \ref{probability}(b1) and (b1').

\begin{figure}[htbp]
\includegraphics*[width=0.47\textwidth]{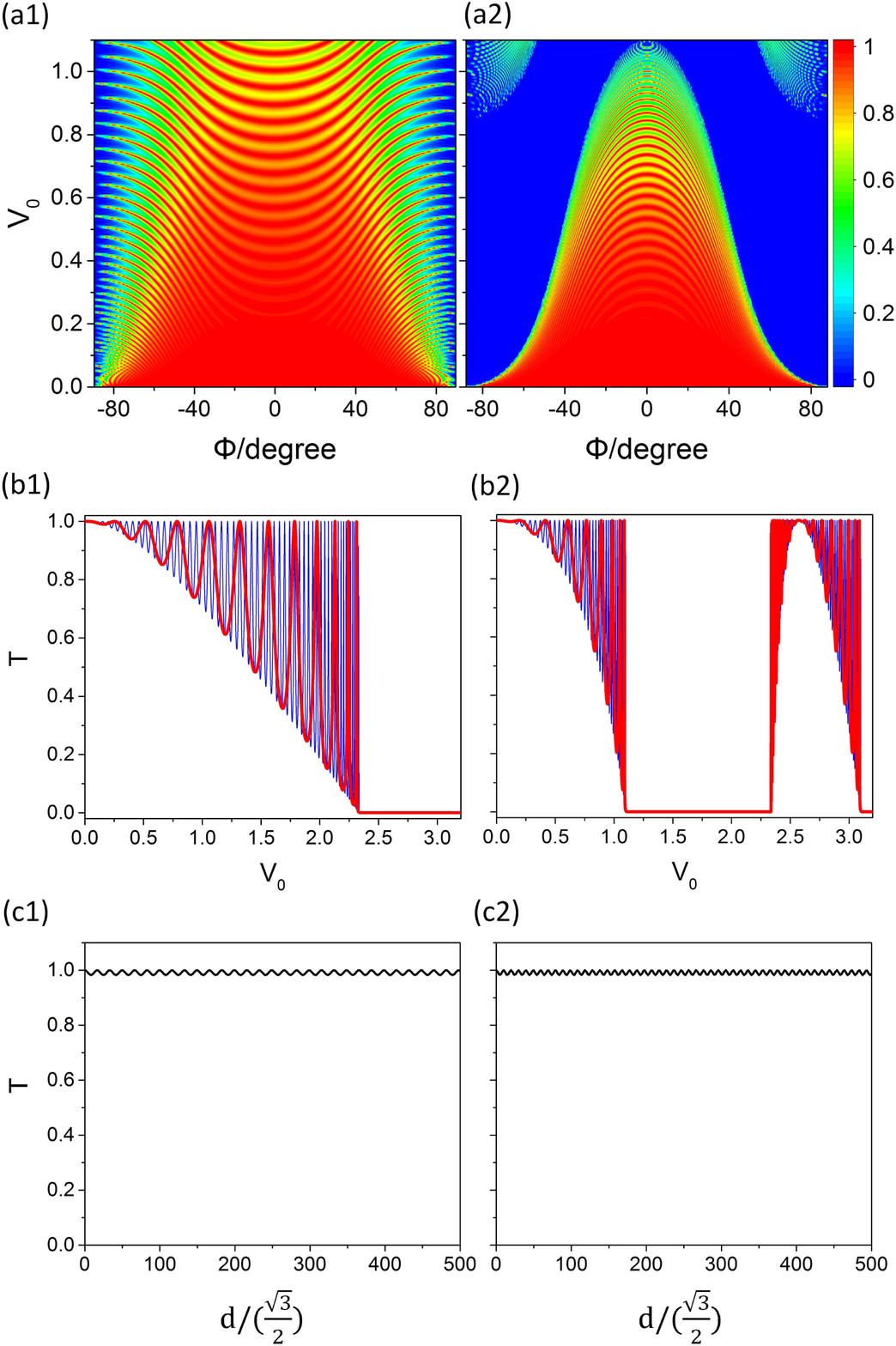}
\caption{(Color online) Total transmissions from the incident state $C$ (left column) and $A$ (right column), for the in-plane incidence $\theta=0$ with Fermi energy $E=0.1$: (a) As functions of the incident azimuthal angle $\phi$ and the barrier potential $V_{0}$, with barrier width $d=100\sqrt{3}/2$; (b) As functions of the barrier potential $V_{0}$, with $d=100\sqrt{3}/2$ (blue) and $d=20\sqrt{3}/2$ (red); (c) As functions of the barrier width $d$, with barrier potential $V_{0}=0.2$. (b) and (c) are for the case of normal incidence with azimuthal angle $\phi=0$.}
\label{v0phi}
\end{figure}

In Fig. \ref{probability}(a1) and (b1),  for both incident electrons from state $C$ and state $A$, the transmission is practically perfect for a large range of angels near normal incidence $0^{\circ}$. The imperfect transmissions at large incident angles have been explained by using Eq. (\ref{T2}) in Sec. III, which also gives the oscillations because $k_{x}$ and $k_{x}'$ are closely related to $k_{y}$, i.e., the angle $\phi$.

For the incidence from state $A$ [Fig. \ref{probability}(b1-b4)], a remarkable feature is that beyond a critical angle, the electron will be completely reflected. This is also a consequence of pseudo-spin conservation in the process of scattering, which is demonstrated in Fig. \ref{electronstates}(b).
For the injecting state $A$, it is on the outer circle, or, on the blue branch with pseudo-spin $\sigma_x \!=\!-1$ [left panel of Fig.\ref{electronstates} (b)]. In the barrier region, the electron has to be scattered onto states with the same $E$, $k_y$ and $\sigma_x \!=\!-1$. Due to the energy shift $V_0>0$ from the barrier potential in this region, these states (blue branch with $\sigma_x \!=\!-1$) are on the inner circle now [middle panel of Fig.\ref{electronstates} (b)]. Therefore, when the injecting angle $\phi$ of state $A$ is large enough on the outer circle, the corresponding $k_y$ in the barrier region will exceed the radius of the inner circle and there will be no states for the electron to be scattered onto. As a result, the injecting electron has to be completely reflected back. From Fig. \ref{electronstates}, it can be seen that with a higher barrier potential $V_0$, the inner circle in the barrier region [the middle panel of Fig. \ref{electronstates}(b)] will be smaller, which leads to a narrower region of $\phi$ that can transport state $A$. This will be seen in Fig. \ref{v0phi}(a2) in the following. For an incidence from state $C$ (on the red branch with $\sigma_x\!=\!+1$), on the other hand, it is a scattering process from an inner circle to an outer circle, so there will always be states in the barrier region to carry the electron, as shown in Fig. \ref{electronstates}(b) and Fig. \ref{v0phi}(a1).

The total transmission probabilities $T_n\equiv \sum_{m} T_{mn}$ (where $m,n\in \{A,C\}$) for a given state  $n$  are shown in Fig. \ref{v0phi}. In the $\phi-V_0$ contours in Fig. \ref{v0phi}(a), two features have been observed in Fig. \ref{probability} and discussed above: almost perfect transmission ($\big|T\big|\sim 1$ with red color) for both states around small $\phi$ and $V_0$, and complete reflections ($T=0$ with blue color) for state $A$ at large $\phi$. Besides these, another prominent feature is oscillating resonances along the direction of $V_0$, which can also be seen in Fig. \ref{v0phi}(b). Most behaviors of these resonances are well consistent with the analytical predictions from the effective model in Section III. During each oscillating period, the maximum total transmissions always recover to unity, as predicted from Eq. (\ref{T2}) in Section III. This reflects the robust nature of the nodal-line semimetals. In the small $V_0$ limit, the oscillating period is nearly independent of $V_0$, consistent with Eq. (\ref{EqResonantCondition}). This equation also predicts that the resonance period should be inversely proportional to the barrier width $d$, which can be verified by comparing the blue and red curves in Fig. \ref{v0phi}(b), corresponding to $d=100\sqrt{3}/2$ and $d=20\sqrt{3}/2$ respectively. The resonant oscillations of the transmission survive until a sudden drop to zero at large $\big|V_0\big|$. This is a trivial consequence of the termination of the carrier band (completely out of the range of all bands or encountering another band with opposite pseudo-spin) due to the energy shift $V_0$ in the barrier region, as can be seen by comparing Fig. \ref{v0phi}(b) with the energy band structure in Fig. \ref{BANDSTRUCTURE}(e).

In Fig. \ref{v0phi}(c), the effect from the barrier width $d$ at normal incidence offers another striking evidence of the extremely robust transport: the transmissions remain perfect (except small fluctuations around $1\%$) even after traversing a barrier range of hundreds of lattice constants. This is similar to the case of single layer graphene (2D Dirac fermion), but drastically contrary to the cases of bilayer graphene or a non-chiral zero-gap semiconductor\cite{Katsnelson2006}.

\begin{figure}[htbp]
\includegraphics*[width=0.47\textwidth]{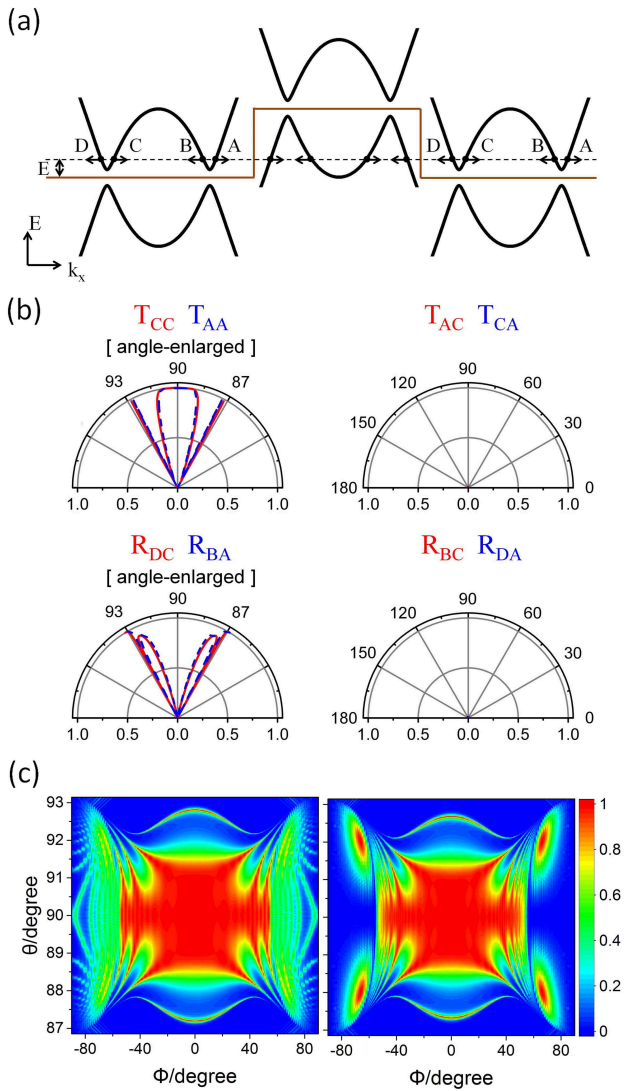}
\caption{(Color online) (a) The schematic of electronic states in three regions of tunneling, when $\theta\!\neq\! 90^{\circ}$. (b) Transmission ($T$) and reflection ($R$) probabilities as functions of the polar angle $\theta$, with $\phi=0$. The red solid (blue dashed) line represents the incidence from state $C$ ($A$). In the first and the third panels, the angle is 10-times enlarged to display the details. (c) The total transmission probability as a function of angle $\phi$ and angle $\theta$ for electron incident from state $C$ (left) and state $A$ (right). The incident Fermi energy $E=0.1$, the potential barrier $V_{0}=0.2$, and the barrier width $d=100\sqrt{3}/2$.}
\label{Ttheta}
\end{figure}

So far in this section, the injecting momentum has been limited to the case of $\theta=90^{\circ}$, i.e., the in-plane injection with $k_z=k\cos(\theta)=0$. Now we investigate the $\theta$-dependence of barrier tunneling around this loop plane. As predicted in last section, there will be two important features in this case: (i) Remarkable transmission can only happen for small $|\theta-90^{\circ}|$ due to the band structure [see Fig. \ref{bands-theta}(b)]; (ii) The scattering process prohibited in the in-plane case may happen since their pseudo-spin orientations are not antiparallel. In Fig. \ref{Ttheta}(b), we plot the $\theta$ dependence of scattering amplitudes among states as labelled in Fig. \ref{Ttheta}(a), with constant $\phi=0$. Indeed, nonzero transmissions or reflections only appear in a very small range of normal incidence: i.e., with $\theta$ (less than $\pm 3^{\circ}$). This feature is also manifested in Fig. \ref{Ttheta}(c), the total transmissions' dependence on $\phi$ and $\theta$.  Notice the $\theta$ scales in the first and the third panels of in Fig. \ref{Ttheta}(b) [and also in Fig. \ref{Ttheta}(c)] have been extremely enlarged to stress the details. Apart from zero angle, the electronic transmissions decay rapidly with fast oscillations, and the electron is reflected as i.e., $C\rightarrow D$ and $A \rightarrow B$. These reflections cannot happen in the case of in-plane injection as shown in Fig. \ref{probability}, since they are on different branches there. In Fig. \ref{Ttheta}(c), the transmissions have a resonant behavior in the direction of $\phi$, which has been predicted in Eq. (\ref{EqResonanceFinitekz}) in last section.

\section{VII. Summary}

In summary, we investigated the barrier tunneling of the loop-nodal semimetal in a hyperhoneycomb lattice by using both analytical and numerical methods. In the case of in-plane incidence (the momentum of the injecting electron in the plane of the loop), most of the scattering behavior for the states near the loop can be understood in the regime of a Klein-type tunneling of a momentum dependent massless 1D Dirac model. In the presence of a low barrier potential, the electron can be almost perfectly transmitted through the rectangular barrier over a wide range of the incident azimuthal angle $\phi$.
When the angle continues to increase, the transmission shows a resonant behavior with respect to the potential height or the angle, with the peak back to perfect transmission almost independent of the barrier width. When the transmission is not perfect, the electrons can only be reflected to the states on the same branch, due to the exact parallel or antiparallel of the corresponding pseudo-spins.

On the other hand, only a small portion of out-of-plane incidents contribute to the transmission, which are limited within a narrow range of the polar angle $\theta$, due to the absence of available states along otherwise directions. Furthermore, in the presence of a mass term, the pseudo-spins between states are not exactly parallel or antiparallel, the reflections will happen between states with different branches, which is called chiral scattering.

\section{Acknowledgements}

This work was supported by National Natural Science
Foundation of China under Grant Nos. 61427901, 11374294 and 11774336, and the Ministry
of Science and Technology of China 973 program under No. 2013CB933304.

\end{document}